\documentclass[11pt,aps,superscriptaddress]{article}

\usepackage{graphicx,graphics,subfigure}
\usepackage{amsfonts,amsmath,amssymb,mathrsfs,bm,amsthm,nccmath,ulem}

\usepackage{array}
\usepackage{tabu}
\usepackage{dcolumn}
\usepackage{float}
\usepackage{url}
\usepackage{ragged2e}
\usepackage{mathtools} 
\usepackage{multirow} 
\usepackage{slashed}
\usepackage[usenames,dvipsnames]{color}
\usepackage{soul}
\usepackage{url}
\usepackage[colorlinks = true,
            linkcolor = blue,
            urlcolor  = blue,
            citecolor = blue,
            anchorcolor = blue,linktocpage]{hyperref}
\usepackage{physics}
\usepackage{makecell}
\usepackage{units}
\usepackage{upgreek}

\usepackage[total={6.5in,9in},top=1in,headsep=0.1in,headheight=1in]{geometry}
\usepackage[utf8]{inputenc}
\usepackage{charter}
\usepackage{fullpage}
\usepackage{hyperref}

\def\bea{\begin{eqnarray}}
\def\eea{\end{eqnarray}}

\definecolor{Zpurple}{RGB}{119, 50, 168}

\definecolor{cardinal}{RGB}{140, 21, 21}

\newcommand\snowmass{
\begin{center}
  \rule[-0.2in]{\hsize}{0.01in}\\
  \rule{\hsize}{0.01in}\\
  \vskip 0.1in
  Submitted to the Proceedings of the US Community Study\\
  on the Future of Particle Physics (Snowmass 2021)\\
  \rule{\hsize}{0.01in}\\
  \rule[+0.2in]{\hsize}{0.01in}\\[-2em]
\end{center}
}

\usepackage[firstpage=true]{background}
\backgroundsetup{contents={\parbox{6.5in}{\snowmass}}, scale=1,placement=top,opacity=1,color=black,position={3.25in,1.2in}}

\usepackage{fancyhdr}
\fancypagestyle{plain}{%
  \fancyhf{}%
  \fancyhead[C]{}
  \fancyfoot[C]{\thepage}
}

\usepackage{authblk}

\usepackage{lipsum}

\title{Snowmass2021 - White Paper\\
Implications of Energy Peak
for Collider Phenomenology: \\
Top Quark Mass Determination and Beyond}
\date{}

\author[1]{Kaustubh Agashe}
\affil[1]{Maryland Center for Fundamental Physics, Department of Physics, University of Maryland, College Park, MD 20742, USA}

\author[1]{Sagar Airen}

\author[2]{Roberto Franceschini}
\affil[2]{Universit\'a degli Studi and INFN Roma Tre, Via della Vasca Navale 84, I-00146, Rome, Italy}

\author[3]{Doojin Kim}
\affil[3]{Mitchell Institute for Fundamental Physics and Astronomy,
Department of Physics and Astronomy, Texas A\&M University, College Station, TX 77845, USA}

\author[1]{Deepak Sathyan}

\begin{document}

\maketitle

\begin{abstract}

We first review the decade-old, broad collider physics research program dubbed energy-peaks. We consider the energy distribution of a massless particle in the lab frame arising from the two-body decay of a heavy particle produced unpolarized, whose boost distribution is arbitrary. Remarkably, the location of the peak of this child particle's energy distribution is identical to its single-valued energy in the rest frame of the parent, which is a function of the parent's mass and that of the other decay product. We summarize generalizations to other types of decay and a variety of applications to BSM. The energy-peak idea can also furnish a measurement of the top quark via the energy of the bottom quark from its decay, which, based on the ``parent-boost-invariance,'' is less sensitive to details of the production mechanism of the top quark (cf.~most other methods assume purely SM production of the top quarks, hence are subject to uncertainties therein, including a possible BSM contribution). The original proposal along this line was to simply use the $b$-jet energy as a very good approximation to the bottom quark energy. This method has been successfully implemented by the CMS collaboration. However, the $b$-jet energy-peak method is afflicted by the jet-energy scale (JES) uncertainty. Fortunately, this drawback can be circumvented by using the decay length of a $B$-hadron contained in the $b$-jet as a proxy for the bottom quark energy. An interesting proposal is to then appropriately dovetail the above two ideas resulting in a ''best of both worlds'' determination of the top quark mass, i.e., based on a measurement of the $B$-hadron decay length, but improved by the  energy-peak concept: this would be free of JES uncertainty and largely independent of the top quark production model. We summarize here the results of such an analysis which is to appear in a forthcoming paper.
\end{abstract}

\clearpage

\tableofcontents

\section{Executive Summary}
It is very well-known that in the rest frame of a parent particle undergoing a two-body decay the energy of each of the child particles is fixed and the expression is defined in terms of parent and the child particle masses.
This implies that we can determine the mass of the parent particle, if we can measure these rest-frame energies of the child particles and if we know the masses of the child particles from other independent measurements. 

However, the parent particle is often created in the laboratory with a boost whose magnitude and direction are a priori unknown. Moreover, boosts of parent particles produced at hadron colliders vary event-to-event. Such a boost distribution depends on the production mechanism of the particle and on the structure functions of the hadrons, and is thus a complicated function. This implies that the energy of the two-body decay product observed in the laboratory frame develops a distribution. Thus it seems like the information that was encoded in the rest-frame energy is lost, and we are prevented from extracting (at least at an easily tractable level) the mass of the parent particle along the lines described above.

Remarkably, it was shown that if one of the child particles from the two-body decay is massless and the parent is unpolarized (i.e., not just restricted to the scalar parent) but no further model assumptions are imposed, then such is not the case: specifically, the distribution of the
child particle’s energy in the laboratory frame has a {\it peak} precisely at its corresponding rest-frame energy~\cite{Agashe:2012bn}. In other words, the peak position is boost-{\it invariant} irrespective of the production details of the parent particle.
In this White Paper, we discuss the idea of energy peak (also called energy-peak method), summarizing its collider applications, especially in terms of mass determinations of the Standard Model (SM) top quark and property measurements of new physics particles. 
We further discuss future directions with the energy peak in the energy-frontier physics program, focusing on a forthcoming study about a new method for the measurement of the top quark mass.  

\section{``Invariance'' of the energy peak} 
We consider the decay of a heavy particle $B$ of mass
$m_B$, i.e., $B \to A +a$ with $A$ having mass $m_A$ and $a$ being a massless visible particle, in which the energy of $a$ in the rest frame of $B$, $E^*$ is given by 
\begin{equation}
E^*=\frac{m_B^2-m_A^2}{2m_B}.
\end{equation}
If parent particle $B$ is boosted by a Lorentz factor $\gamma$ in going to the laboratory frame, then the $a$ energy seen in the laboratory
frame $E$ is 
\begin{equation}
    E=E^*\gamma(1+\beta \cos\theta^*)\,,
\end{equation}
where $\theta^*$ defines the direction of emission of particle $a$ in the $B$ rest frame with respect to the boost direction $\vec{\beta}$ of $B$ in the laboratory frame.  Due to our assumption of the parent being not polarized, the probability distribution of $\cos\theta^*$ is flat. This implies that, for a fixed $\gamma$, the distribution of $E$ is flat as well. More precisely, since $\cos\theta^* \in [-1, 1]$, for any fixed $\gamma$ the shape of the $E$ distribution is a simple ``rectangle'' spanning the range
\begin{equation}
    E \in \left[E^*\left(\gamma-\sqrt{\gamma^2-1}\right), E^*\left(\gamma+\sqrt{\gamma^2-1}\right) \right]\,. \label{eq:range}
\end{equation}
A few crucial observations are in order. First, the lower (upper) bound of \autoref{eq:range} is smaller (larger) than $E^*$ for an arbitrary $\gamma$, which implies that every rectangle contains $E^*$. Remarkably, $E^*$ is the only energy value to enjoy such a property as long as the distribution of parent particle boost is non-vanishing in a small region around $\gamma=1$. Furthermore, the energy distribution being flat for every $\gamma$, there is no other value of the energy which gets a larger contribution than $E^*$. Thus, upon ``stacking up'' the rectangles of different widths, corresponding to a range of $\gamma$'s, we see that the peak of the $a$ energy distribution is unambiguously located at $E=E^*$~\cite{Agashe:2012bn}. We emphasize that no model details other than a {\it two-body} decay of {\it unpolarized} parent $B$ into a {\it massless} child $a$ are assumed, i.e., this ``{\it invariance}'' property against the boost is {\it model-independently} valid. 

More formally, the $a$ energy spectrum $f(E)$ is given by superimposing the above-mentioned rectangles weighted by the boost distribution $g(\gamma)$:
\begin{equation}
    f(E)=\int_{\frac{1}{2}\left(\frac{E}{E^*}+\frac{E^*}{E} \right)}^\infty d\gamma \frac{g(\gamma)}{2E^*\sqrt{\gamma^2-1}}\,. \label{eq:int}
\end{equation}
Since $g(\gamma)$ is in general unknown, a closed form of the above expression is generally not available. One may instead employ an ansatz to capture the functional properties encoded in \autoref{eq:int}: $i$) $f(E/E^*)=f(E^*/E)$; $ii$) $f_{\max}=f(E=E^*)$; $iii$) $f(E\to 0,\infty)\to 0$; and $iv$) $f(E)\to \sim \delta(E-E^*)$ in some limit of its parameters. These properties are not sufficient to single out a functional form for $f(E)$, so we proposed a successful ansatz of the following form~\cite{Agashe:2012bn}
\begin{equation}
    f(E) \propto \exp\left[-\frac{p}{2}\left(\frac{E}{E^*}+\frac{E^*}{E} \right) \right],
\label{eq:ansatz}
\end{equation}
where $p$ is a parameter which encodes the width of the peak.

\section{Applications}

In this section, we summarize the physics applications of the energy-peak observation, beginning with the top quark mass measurement followed by new particle property measurements.

\subsection{Top quark mass measurement}
The idea of energy peak was first applied to the top quark mass measurement, as a top quark (identified as $B$) is produced in an {\it unpolarized} way due to QCD interactions and decays to {\it two} child particles, a $W$ gauge boson (identified as $A$) and a bottom quark (identified as $a$)~\cite{Agashe:2012bn}. 
Note that the method is not only insensitive to the production details of top quark such as QCD effects, PDF uncertainties etc but valid even in the case where production is ``contaminated'' by non-SM contributions as far as unpolarized production of top quarks holds, whereas many of other traditional methods rely on the assumption of SM top quark. 
Here the bottom quark is not massless, but it is so boosted in the top quark decay that the phase space invalidating the above argument is negligible. 
We studied a detector-level sample of fully leptonic top decays from the process, $pp \to t\bar{t}\to b\bar{b}\mu^-e^+\nu_e\bar{\nu}_\mu$, at the LHC7 with an ansatz of the $b$-jet energy spectrum shown in \autoref{eq:ansatz}, and demonstrated that the extracted mass is almost the same as the input value with an error of $\sim 1.5\%$. 

The CMS Collaboration adopted this idea to perform a (complementary) measurement of the top quark mass using the $b$-jet energy distribution and reported $m_t=172.29\pm1.17 ({\rm stat.})\pm 2.66 ({\rm syst.})$~\cite{CMS:2015jwa}. One of the dominant sources of systematics is the one from jet energy scaling which would be mitigated with higher statistics, so the upcoming (high-luminosity) LHC would provide a better opportunity for a more precise measurement along the line. 

Indeed, understanding higher-order effects, especially final state radiation of $b$ quark, is crucial to reduce the systematics in the top quark mass measurement. In a follow-up work~\cite{Agashe:2016bok}, we investigated our $b$-jet energy method with QCD NLO taken into account at the LHC14, and showed that for a 1\% jet energy scale uncertainty the systematic error estimate would be improved to $\pm (1.2({\rm exp})+0.6 ({\rm th}))$~GeV. As an alternative route to get around the jet energy scaling issue, we studied the idea of using the $B$-hadron energy which can be measured at the tracker and ECAL~\cite{Agashe:2016bok}. 
In a similar spirit, we further investigated various $B$-hadron observables again at the LHC14 and pointed out that constraining the relevant Monte Carlo parameters in event generators such as \textsc{Pyhia 8} and \textsc{Herwig 6} by $\mathcal{O}(1-10\%)$ would allow us to determine the top quark mass within an uncertainty of $\lesssim0.5$~GeV~\cite{Corcella:2017rpt}.
We believe that these ideas can be tested in the upcoming LHC runs and make a contribution to the task of precision top quark mass measurements. 

\subsection{New physics} 
Applications of the energy-peak method are not limited to the SM top quark mass measurement, but readily extended to property measurements of new physics particles. In particular, since the method is valid irrespective of visibility of the other decay product, it can be useful for a wide range of new physics models including the ones containing dark matter candidate(s). 
To show its broad applicability in realistic examples, we performed a few benchmark studies in the scenario where pair-produced parent particles undergo a two-step cascade decay terminating in an invisible particle which arises in many of the new physics models~\cite{Agashe:2013eba}, in the scenario where a parent particle goes through a three-body decay~\cite{Agashe:2015wwa}, and in the scenario where the decay products are (non-negligibly) massive~\cite{Agashe:2015ike}. 
Beyond the mass measurements of new particles, the energy-peak method can be used for distinguishing dark matter stabilization symmetries, e.g., $Z_2$ vs. $Z_3$, in combination with the $M_{T2}$ variable~\cite{Agashe:2012fs}. 

These studies essentially cover most of the scenarios that arise in typical new physics models such as supersymmetry, extra dimensions, and dark matter models. Therefore, upon discovery of new physics at the upcoming LHC or future colliders, we expect that the energy-peak method can play an important role in unmasking the underlying model details.   

\section{Forthcoming work} 
We have proposed a new top quark mass measurement technique, developing a method in which the energy-peak results is infused into $B$-hadron decay length predictions, hence enabling a more robust  extraction of the top quark mass from length measurements.
The $B$-hadron decay length can be considered as a ``proxy'' for its energy. 
Similarly to the aforementioned technique of utilizing $B$-hadron observables, we expect that the method will be unaffected by the jet energy scale uncertainty as the main observable is the decay length of (long-lived) $B$-hadrons measured in the tracker of LHC detectors.
Indeed, a similar technique was used to measure the top quark mass by the CMS Collaboration with the assumption of the  top quark kinematics well predicted by the SM~\cite{CMS:2013cea}. 
By contrast, we expect that the involvement of the energy peak will ensure better model-independency in the extraction of the top quark mass due to  {the insensitivity of energy peaks to production mode dynamics.} 
{In particular, we expect a reduced sensitivity to top quark transverse momentum spectrum, which was found to have a large impact on the CMS result.}
{The flip side of this proposal is that good knowledge of hadronization is necessary to carry out the measurements. This is the case in general for measurements based on hadron-level observables including  the aforementioned CMS measurement~\cite{CMS:2013cea}.}

\subsection{Decay length method: formalism} 
In the following we describe  schematically the general idea of the method, for the actual implementation the reader is referred to \cite{inprogres}.
In general, the $B$-hadron decay length ($L_B$) distribution, denoted by $G \left( L_B \right)$, can be obtained as follows. We start with the distribution of the energy of bottom quark in the laboratory frame, denoted by $f \left( E_b \right)$; convoluting it first with fragmentation function $D$ in order to go to the $B$-hadron energy;
then relating this energy to the $B$-hadron mean decay lifetime, $\tau^{\rm rest}_{B}$, and finally, 
a second convolution with the usual decay exponential to obtain the length. This procedure gives:
\begin{eqnarray}
G \left( L_B \right) 
& = & \int d E_B \int d E_b \; 
f \left( E_b \right) 
D \left( \frac{ E_B }{ E_b }; E_b \right) 
\frac{ m_B }{ c \tau_B^{ \rm rest } E_B }
\exp \left( - \frac{ L_B m_B }{ c \tau_B^{ \rm rest } E_B } \right) \,,
\label{eq:LB-general}
\end{eqnarray}
where $f \left( E_b \right)$ depends on
top quark mass $m_t$
 and the boost distribution of top quark itself in the laboratory frame, which in turn
depends on the details of its production mechanism.

The above formula written in a schematic way allows to easily explain our idea and differences with other methods. Indeed we have written the above expression with $\beta\approx 1$ as $B$ hadrons are ultra relativistic. In our actual analysis,  we use simulated data from MadGraph5\cite{Alwall:2014hca}  and Pythia8\cite{Sjostrand:2014zea} throughout, hence we use fully realistic data and templates predictions without simplifying assumption, e.g.  the actual velocity of hadrons is always considered.

{Owing to the generality of \autoref{eq:LB-general}, it  can be specialized to the  CMS $B$-hadron decay length method.} This method assumes that the top quarks are produced by the SM matrix element, therefore {the prediction for the decay length distribution} can be written as: 
\begin{eqnarray}
G^{ \rm fit, \; SM } \left( L_B; m_t \right) & \approx & \int d E_B \int d E_b  f^{ \rm SM } \left( E_b; m_t \right)
D \left( \frac{ E_B }{ E_b }; E_b \right) 
\frac{ m_B }{ c \tau_B^{ \rm rest } E_B }
\exp \left( - \frac{ L_B m_B }{ c \tau_B^{ \rm rest } E_B } \right) \;,
\label{eq:LB-SM}
\end{eqnarray}
where we denoted the energy distrubtion with the label ``fit'' to stress that it will later be used in a template fit, $f^{ \rm SM } \left( E_b \right)$ denotes the boost distribution of top quarks predicted in the SM under suitable approximations adopted in the experimental analysis {(e.g., NNLO in QCD perturbation theory, with or without parton shower, etc)}.
{As theoretical predictions can be computed for any mass,  $m_t$ is taken to be a parameter and can be  fitted from the observed data.} %
For the case of the  CMS measurement~\cite{CMS:2013cea}, in the results shown in the following, the above equation is specialized to the 
transverse component of the decay length denoted by $L_{ x y }$. 

Inspired by earlier results on  energy distributions around their peak, (see \autoref{eq:ansatz}), we use a modified ansatz
\begin{equation}
f \left( E_b \right)= \exp \Big[ - w  \left( \frac{ E_b }{ E_b^{\rm rest} } +  
\frac{ E_b^{\rm  rest } }{ E_b } \right)^{\nu} \Big]\,,
\end{equation}
so that we propose a template fit as follows:
\begin{eqnarray}
G^{ \rm fit, \; us } \left( L_B; E^{ \rm rest }_b, w \right) & \approx & \int d E_B \int d E_b  \frac{1}{ N (w) } \exp \Big[ - w  \left( \frac{ E_b }{ E_b^{\rm rest} } +  
\frac{ E_b^{\rm  rest } }{ E_b } \right)^\nu \Big] \times \nonumber \\
& & 
D \left( \frac{ E_B }{ E_b }; E_b \right) 
\frac{ m_B }{ c \tau_B^{ \rm rest } E_B }
\exp \left( - \frac{ L_B m_B }{ c \tau_B^{ \rm rest } E_B } \right) \,.
\label{eq:LB-us}
\end{eqnarray}
Here $N(w)$ is a normalization factor, and we pick $\nu=0.3$ fixed from optimization of the overall result, while  
$E^{ \rm rest }_b$ and $w$ are fit parameters. In particular, from  the extracted value of $E^{ \rm rest }_b$ we can get  $m_t$ using the relation:
\begin{eqnarray}
E^{ \rm rest }_b & = & \frac{ m_t^2 - m_W^2 + m_b^2 }{ 2 m_t }\,.
\label{eq:Eb-rest}
\end{eqnarray}

\subsection{Test of the production mode sensitivity of the top quark mass measurement}
We can then test the model-independence of our method  and compare with  the CMS $L_{ xy }$ method as follows. We produce pseudo-data using the 
SM Monte Carlo predictions done with MadGraph5\cite{Alwall:2014hca}  and Pythia8\cite{Sjostrand:2014zea}.  In our analysis we used the actual length distribution predicted by these Monte Carlo packages for some choices of top quark mass, denoted by $m_t^{ \rm input }$. This prediction can be thought schematically as arising from
\begin{eqnarray}
G^{ \rm (simulated) \; data } \left( L_B; m_t^{ \rm input } \right) & \approx & \int d E_B \int d E_b \; f^{ \rm SM } \left( E_b; m_t^{ \rm input } \right) \nonumber \\
&\times& D \left( \frac{ E_B }{ E_b }; E_b \right) 
\frac{ m_B }{ c \tau_B^{ \rm rest } E_B }
\exp \left( - \frac{ L_B m_B }{ c \tau_B^{ \rm rest } E_B } \right) \,,
\label{eq:LB-data}
\end{eqnarray}
 where $f^{ \rm SM } \left( E_b; m_t^{ \rm input } \right)$ is taken as the MadGraph5\cite{Alwall:2014hca}  and Pythia8\cite{Sjostrand:2014zea} computation. 
 
 {Hadronization functions $D$ are in principle a necessary input to this calculation. 
 In our simplified analysis, we assume that these are perfectly known from theory and experiments on hadronization of $B$-hadrons. The effects of the imperfect knowledge of these functions will not be explored in this white paper. In practice we generate pseudo-data from Pythia8 and we ``measure'' hadronization functions in the Pythia8 runs. The ``measured'' hadronization functions are used directly in our numerical computations of integrals thus eliminating any inaccuracy due to lack of knowledge of hadronization effects.
 The effects of possible imperfect knowledge of hadronization functions will be presented elsewhere~\cite{inprogres}.} 
 {Bottom quarks can decay into many different $B$-hadrons, whereas  \autoref{eq:LB-data} describes only one species of B hadrons. Hence, the rest frame lifetime, mass, and relative fraction of all the $B$-hadrons species are crucial inputs, and a generalized function is used. For the analysis presented in this white paper, we assume everything is known perfectly. The sensitivity to this knowledge will be presented elsewhere~\cite{inprogres}}.

In order to extract a best-fit $m_{t}$ we perform a fit to the pseudo-data \autoref{eq:LB-data}   for each of the two predictions described around \autoref{eq:LB-SM} and \autoref{eq:LB-us}.
To mimic a realistic measurement, selection cuts are imposed to eliminate background, which could lead to bias in the best-fit $m_t$ when using any of the two methods without any modification. This effect can be very easily mitigated for the function in \autoref{eq:LB-SM}, by simulating the same effect in the templates. For \autoref{eq:LB-us}, we tune the lower and upper bound on the $b$-quark integral in a way that the bias generated is compensated for. We chose cuts very similar to the ones ATLAS and CMS Collaborations use (see e.g.~\cite{CMS:2021vhb}), with all the transverse momentum cuts set to $25$ GeV for simplicity. The lower and upper bounds on $E_b$ are $40$ GeV and $450$ GeV, respectively. 

Of course both approaches should ideally (neglecting statistical error or those due to fragmentation functions etc.) give correct values in this situation. This is obvious for SM fit described around \autoref{eq:LB-SM}, but it is not obvious for our method based on templates of the form
\autoref{eq:LB-us} as we have used in it an ansatz for the energy distribution \autoref{eq:ansatz}. Performing the extraction of $m_{t}$ using  templates based on  \autoref{eq:LB-us}, we obtain a faithful extraction, which is not surprising, given that \autoref{eq:ansatz} has already been successfully used in previous works, e.g., Ref.~\cite{Agashe:2012bn}.

In order to test the robustness of our method and compare with the methods based on hard-wired SM predictions of energy distribution we repeat the fits using ``tweaked'' data. The tweak we adopt aims at testing the robustness of the mass measurement methods upon changes of the top quark production kinematics, e.g. its transverse momentum.  This can be thought of as a modified boost distribution $f^{SM}$, which mimics a mismatch between the actual boost distribution in data and most precise theory predictions which could be due to BSM contribution in top quark production or simply due to effects in the SM that are not correctly captured in the most precise theory predictions.
Denoting the corresponding tweaked bottom quark energy distribution by $f^{ \widetilde{ \rm SM } } \left( E_b \right)$, the new pseudo-data can be described  as 
\begin{eqnarray}
\tilde{G}^{ \rm (simulated) \; data } \left( L_B; m_t^{ \rm input } \right) & \approx & \int d E_B \int d E_b \; 
f^{ \widetilde{ \rm SM } } \left( E_b; m_t^{ \rm input } \right) \nonumber \\
&\times& D \left( \frac{ E_B }{ E_b }; E_b \right) 
\frac{ m_B }{ c \tau_B^{ \rm rest } E_B }
\exp \left( - \frac{ L_B m_B }{ c \tau_B^{ \rm rest } E_B } \right) \,.
\label{eq:LB-new-data}
\end{eqnarray}
 In our work we obtain the tweaked  data by reweighting our MC events according to the the $p_{T}$ of the top quark.
The new weight $\tilde{w}$ for each event is obtained as a linear function of the top quark $p_{T}$ according to
\begin{equation}
\frac{\tilde{w} }{w} = 1 + \alpha \cdot \left(   p_{T} - 200\; {\rm GeV} \right) \;{\rm for}\;  p_{T}<400   \; {\rm GeV } \,.
\label{eq:reweight}
\end{equation}

The tweaked distribution that we construct corresponds to the same input top mass, so that we can assess the performance of the method comparing the extracted top quark mass with the original input value.
Following the same approach as before, we fit to the tweaked pseudo-data both templates for the hard-wired SM prediction described around \autoref{eq:LB-SM} and templates from our proposal \autoref{eq:LB-us}. 
Clearly, a method based on a hard-wired SM prediction like \autoref{eq:LB-SM},  even in this ideal setting where hadronization is perfectly known, cannot give us a perfect fit. In  fact, even if the top quark mass value is the same, the shape of $f$ and $\tilde{f}$ distributions entering \autoref{eq:LB-SM} and \autoref{eq:LB-new-data} will manifestly not agree {(albeit the location of their peaks  will be the same given its top quark boost-invariance!)},
hence $\tilde{G}^{ \rm (simulated) \; data }$
cannot be exactly be same as $G^{ \rm fit \; SM }$.
The best fit will
then give $m_t \neq m_t^{ \rm input }$.

Our method, instead, has a chance of not returning a biased $m_{t}$ due to the change of top quark kinematics.
The point is that $f^{ \widetilde{ \rm SM } }$ is also expected of the  form in \autoref{eq:ansatz}, since the latter form is independent of boost distribution of parent top quark.
Therefore, while $m_t \; \hbox{(in  \autoref{eq:LB-us})} = m_t^{ \rm input } \; \hbox{(in \autoref{eq:LB-new-data})} $ makes the peaks agree just like for the SM fit, the choice of width parameter $w$ in \autoref{eq:LB-us}
 gives enough ``flexibility'' to get the remaining parts of the $f \left( E_b \right)$
distributions to also line-up for our fit, 
hence in principle  (i.e., neglecting statistical, hadronization etc. uncertainties.) at least
$\tilde{G}^{ \rm (simulated) \; data }$
 can  be exactly be same as $G^{ \rm fit \; us }$. 
In general, the tweaked data gives rise to a different value of $w$, even if $m_t$ was the same.

\begin{figure}[h]
\centering
\includegraphics[width=0.48 \linewidth
,angle=0]{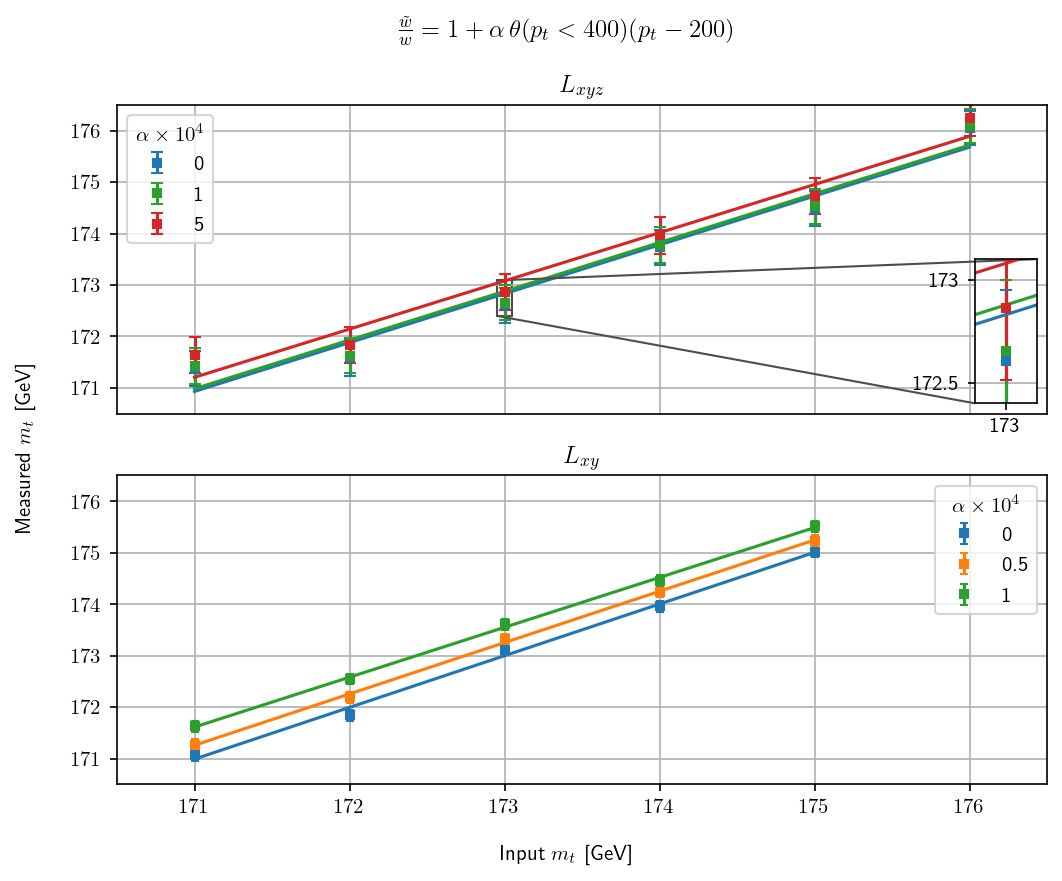}
\includegraphics[width=0.48 \linewidth
,angle=0]{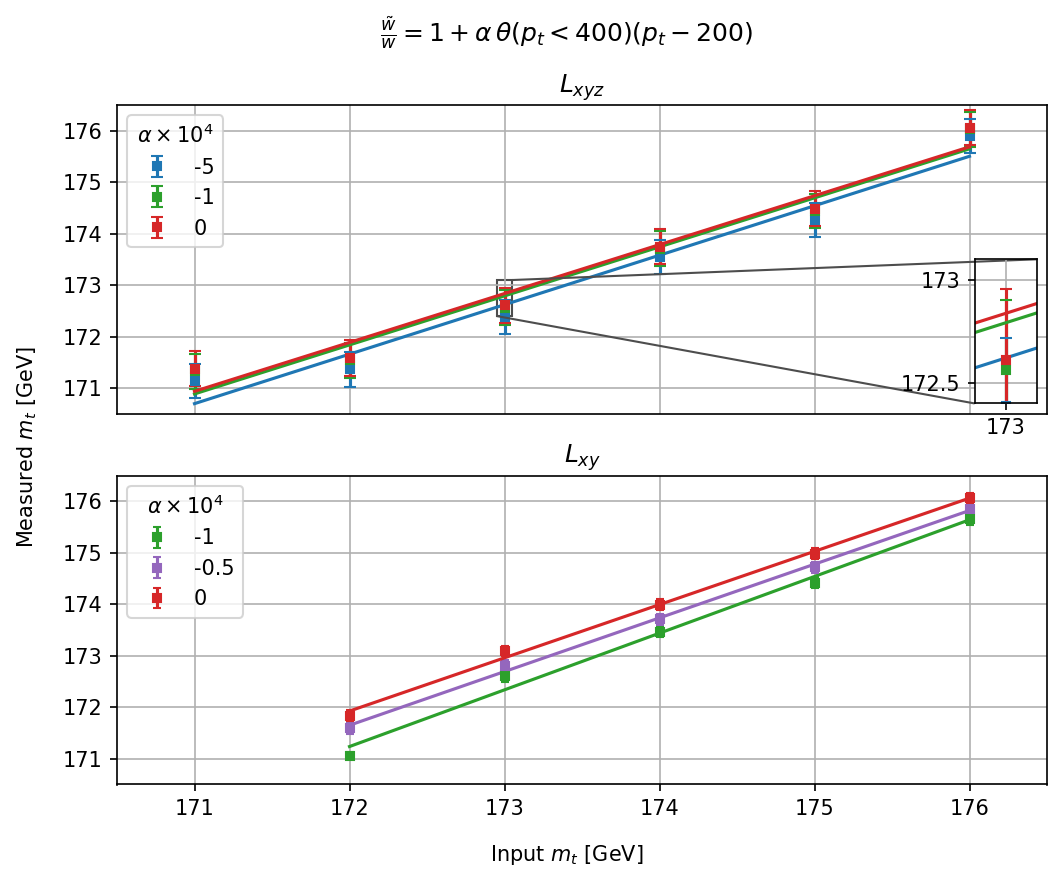}
\caption{Comparison of the extracted top quark mass from pseudo-data with reweighted $p_{T,top}$ kinematics according to \autoref{eq:reweight}. Results for template fitting using the $L_{xy}$ observable with a hard-wired $b$-quark energy distribution (bottom panel) and our method (top panel). Results on the left (right) correspond to hardening (softening)   the $p_{T,top}$ spectrum of the pseudo-data. }
\label{figure:money-us-vs-Lxy}
\end{figure}

This value-added by our method is illustrated in a preliminary result 
\autoref{figure:money-us-vs-Lxy},
where 
the extracted value of $m_t$ is plotted vs.~its input value.
The top panel is for our method, while bottom one is for CMS $L_{ xy }$.
The $\alpha=0$ line is the case of simulated data assuming SM production, where we see that extracted and input values of 
$m_t$ agree quite well for both methods, as expected based on above discussion.
Colored lines in the left (right) part of the plot give the results for different degrees of hardening (softening) the top quark $p_T$ distribution
(and equivalently to that on boost). The change corresponding to  $\alpha = 10^{-4}$ gives a  modification of
the average $p_T$ by $\sim 0.5 \%$, that is roughly the same size of the re-weighting needed to have LHC data to match MC predictions~\cite{CMS:2021vhb}. In left panels this case  is plotted as a green line. The extracted $m_t$ in orange gives the case of ``mid-way'' amount of re-weighting.
We find that the CMS $L_{ xy }$ method extracts a value of $m_t$ which is off from input by $\sim 600$ MeV in this case.
Strikingly, our method has a negligible change in the extracted top quark mass
for such a size of change in top quark $p_T$ distribution, as seen from magnified inset. In fact, we have to increase the re-weighting in
$p_T$ by a factor of 5 (red line) in order to make it ``visible'' on the same scale as bottom panel for CMS method. Similar conclusions hold from the lines in the right panel of \autoref{figure:money-us-vs-Lxy}, which deals with negative $\alpha$ corresponding to softening the pseudo-data.

\medskip

\section{Conclusions} 
In conclusion, the idea of energy peak is not only theoretically interesting {\it per se} but providing a kinematic handle to measure the mass of the SM top quark irrespective of its production details and to extract properties of new physics particles. 
Given the scientific applications of the energy peak method, it will be an important aspect of both the energy-frontier physics program and the theory-frontier physics program in the next decade and beyond. 

\section*{Acknowledgments} 
The work of KA, SA and DS was supported in part by the NSF grant PHY-1914731 and by the Maryland Center for Fundamental Physics. 
The work of DK is supported by the DOE Grant No. DE-SC0010813.


\bibliography{ref}
\bibliographystyle{unsrt}
\end{document}